\documentclass[journal,twoside]{IEEEtran}
\usepackage{cite}       % Written by Donald Arseneau
\usepackage{graphicx}   % Written by David Carlisle and Sebastian Rahtz

\newcommand{\ie}{{\it i.e.\,}}

\newcommand{\vs}{{\it vs.\,}}

\begin{document}
\title{Two tone response in Superconducting Quantum Interference Filters}
\author{P.~Caputo, 
	J.~Tomes, 
	J.~Oppenl{\"a}nder, 
	Ch.~H{\"a}ussler, 
	A.~Friesch, 
	T.~Tr\"auble,
	and N.~Schopohl% <-this % stops a space
%\thanks{Manuscript received August 29, 2006; revised....}%
        %This work was supported by the IEEE.}% <-this % stops a space
\thanks{P.~Caputo, J.~Tomes, J.~Oppenl{\"a}nder, Ch.~H{\"a}ussler, A.~Friesch, and T.~Tr\"auble are with QEST, Quantenelektronische Systeme GmbH, Steig{\"a}ckerstr. 13, 72768 Reutlingen (Germany).}
\thanks{N. Schopohl is with Lehrstuhl f\"ur Theoretische Festk{\"o}rperphysik, Universit{\"a}t T{\"u}bingen, Auf der Morgenstelle 14, 72076 T{\"u}bingen (Germany).} }
% note the % following the last \IEEEmembership and also the first \thanks - 
% instance, "A\textbf{} \textbf{}B" will typeset as "A B" not "AB". If you want
% "AB" then you have to do: "A\textbf{}\textbf{}B"
\markboth{IEEE Transactions on Applied Superconductivity,~Vol.~x, No.~x,~x~x}{Caputo \MakeLowercase{\textit{et al.}}: Two tone response in Superconducting Quantum Interference Filters}
% *** Note that you probably will NOT want to include the author's name in ***
% *** the headers of peer review papers.                                   ***

% If you want to put a publisher's ID mark on the page
% (can leave text blank if you just want to see how the
% text height on the first page will be reduced by IEEE)
%\pubid{0000--0000/00\$00.00~\copyright~2006 IEEE}

%\specialpapernotice{(Invited Paper)}

\maketitle

\begin{abstract}

We successfully exploit the parabolic shape of the \textit{dc} 
voltage output dip around $B=0$ of a Superconducting Quantum Interference Filter (SQIF) to mix weak external \textit{rf} signals. The two tone response of weak time harmonic electromagnetic fields has been detected on the spectral voltage output of the SQIF at frequency $f_0 = f_1 - f_2$, for various frequencies $f_1$ and $f_2$ ranging from few MHz up to 20 GHz. The two tone response is a characteristic function of static magnetic field $B$ and of bias current $I_\mathrm{b}$, related to the second derivative of the \textit{dc} voltage output.

\end{abstract}


% Will be ignored for other modes.
\IEEEpeerreviewmaketitle

\section{Introduction}
% \PARstart{A}{demo} file is ....
% \PARstart{A}{}demo file is ....
% \PARstart{T}{his demo} file is ....


\PARstart{S}{uperconducting} Quantum Interference Filters (SQIFs), when operated in the resistive mode, have been shown to be effective flux-to-voltage transformers with a high transfer factor \cite{Oppenlander:PRB,Haussler:JAP2001} and large voltage swing.
Employing various flux focusing
structures together with a SQIF it is possible to further significantly enhance the
transfer factor \cite{Schultze:SST_2003,Schultze:SST_2005}. 

In the presence of incident time dependent electromagnetic fields, the spectral voltage output $\widehat{V}(I_\mathrm{b}, B,f)$ of a SQIF is the Fourier transform of the voltage
$V(I_\mathrm{b} + i_\mathrm{rf}(t), 
B + b_\mathrm{rf}(t),t)$. 
The dependence of the coupling strength of the incident \textit{rf} signals depends on the bias current $I_\mathrm{b}$ and the static field $B$. It  
can be estimated for slowly varying signals as
\begin{eqnarray}
\label{Eq:Vdc(t)}
&& \left\langle V(\overbrace{I_\mathrm{b} + i_\mathrm{rf}(t)}^{\mathrm {bias \,\, coupling}}, 
\underbrace{B + b_\mathrm{rf}(t)}_{\mathrm {flux \,\, coupling}},t) \right\rangle  = %\nonumber
\\
&=&  V_\mathrm{dc}(I_\mathrm{b}, B)\,+\, (i_\mathrm{rf}(t) \, \partial_\mathrm{I_b} \,+\, b_\mathrm{rf}(t)\, \partial_\mathrm B) V_\mathrm{dc}(I_\mathrm{b}, B)  \nonumber \\ 
  &\,+\,&  \frac12 (i_\mathrm{rf}(t) \, \partial_\mathrm{I_{b}} \,+\, b_\mathrm{rf}(t) \, \partial_B)^2 V_\mathrm{dc}(I_\mathrm{b}, B)  + \ldots \nonumber  
\end{eqnarray}
Here $V_\mathrm{dc}(I_\mathrm{b}, B)$ is the dc voltage output of the SQIF in the absence of any incident \textit{rf} signals; $\partial_\mathrm{I_{b}}$ and $\partial_\mathrm B$ denote the partial derivatives with respect to $I_\mathrm{b}$ and $B$, respectively. The fluctuation $i_\mathrm{rf}(t)$ of the bias current describes the response to the incident electric field $e_\mathrm{rf}(t)$.

In recent work \cite{Caputo:APL_2006}, we investigated the two tone response of the SQIF to weak time harmonic \textit{rf} magnetic fields $%
b_{\rm {rf}}\left( t\right) =b_\mathrm{rf1}\cos \left( 2\pi f_{1}\,t\right) +b_\mathrm{rf2}\cos
\left( 2\pi f_{2}\,t\right)$, in a frequency range up to a few hundred MHz.

In the following we study the response of the SQIF
to weak time harmonic \textit{rf} electromagnetic fields $b_\mathrm{rf}(t)$ and $e_\mathrm{rf}(t)$ consisting of two tones in the frequency range from few MHz up to 20 GHz. 
The two tone response is studied as a function of static magnetic field $B$ and of bias current $I_\mathrm{b}$.

 
%\hfill August 29, 2006

\hfill

\section{Two tone experiments}

\subsection{The setup}

\begin{figure}[!tb]
\centering
\includegraphics*{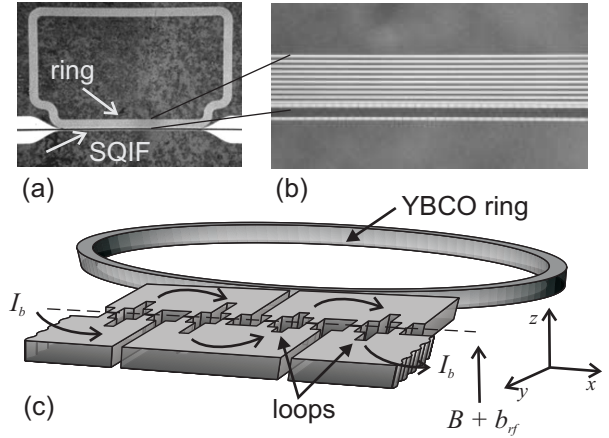} 
\caption{(a) Optical microscope image of the chip with
the SQIF placed along the grain boundary, and the superconducting ring inductively
coupled to the SQIF; (b) zoom on the ring made of 10 equidistant parallel loops, one aligned inside the other. All superconducting
parts, except the regions across the junctions, are covered by gold.
(c) Sketch, not in scale, of the SQIF-sensor; the dashed line represents the grain boundary; the meandered path of the bias current and the direction of the magnetic field are shown by arrows.}
\label{chip}
\end{figure}

Our SQIFs are manufactured from YBa$_{2}$Cu$_{3}$O$_{7-x}$ grain boundary Josephson junctions grown on 24$^{\circ }$-oriented bicrystal
MgO substrates\cite{IPHT_Jena}. The junction width is $2\,\mathrm{\mu m} $, the YBa$_2$Cu$_3 $O$_{7-x}$ layer is  $130\,
\mathrm{nm}$ thick, so that the resulting junction critical current density
is $J_\mathrm{c} \approx 23 \,\mathrm{kA/cm^2} $, at $T=77\,\mathrm{K}$. The 
SQIF consists of 211 loops biased in series, the distribution of the
loop areas ranging between $38\mathrm{\mu m^{2}}$ and $210\,\mathrm{\mu m^{2}%
}$.  
The wirings connecting the loops in series form a meandered path   
across the grain boundary, as sketched in Fig.\thinspace\ref{chip}(c). In order to provide flux focusing, we
fabricated a SQIF with a superconducting ring, placed at one side of the grain boundary and inductively coupled to the SQIF. 
Figure\thinspace\ref{chip}(a) is an optical microscope image of the SQIF-sensor. The focusing loop is effectively a split loop design, consisting of 10 equidistant parallel thin loops, one
aligned inside the other; calculations show \cite{Ludwig:IEEE2001} that the split ring design leads, in average, to a larger current density compared to a single ring of the same cross section. With the split ring, we have achieved an enhanced sensitivity to the magnetic field, \ie a larger transfer factor $V_\mathrm{B} = \mathrm{max}(\partial V/ \partial B)$. % with respect to the case of a single broad ring.
A fragment of the split ring is in Fig.\thinspace\ref{chip}(b).

The static magnetic field $B$ is applied via a
multi-turn coil placed inside the dewar; the \textit{rf} magnetic field $%
b_\mathrm{rf}(t)$ is broadcast by a $50\,\mathrm{\Omega }$-loop antenna (outside the dewar). For the two tone experiments, the \textit{rf} field is a superposition of two time harmonic signals, applied to the ´loop antenna by means of two independent generators. The experiments
presented here were made with a mu-metal shield surrounding the dewar; the loop
antenna was placed inside the shield, at a distance of about $5\,\mathrm{cm}$
from the chip.

\begin{figure}[!tb]
\centering
\includegraphics*{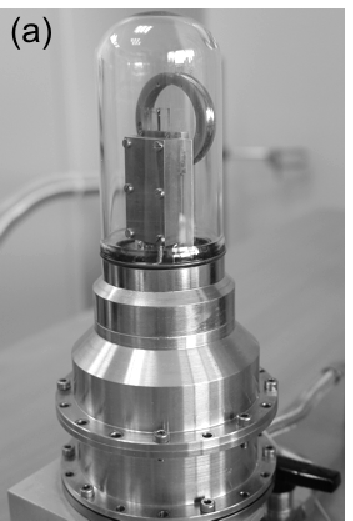} 
\includegraphics*{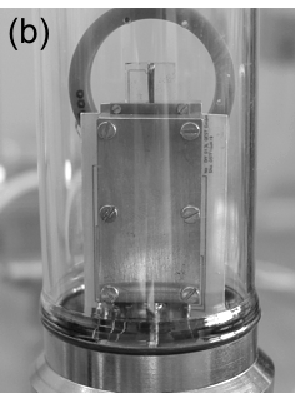} 
\caption{Experimental setup with the cryocooler and the glass dewar (a); enlarged view of the chip and of the amplifier (b). The chip is anchored by one side at the cold finger and coupled to the amplifier by a coplanar line. Directly under the chip, the multi-turn coil is mounted, for the static field $B$.}
\label{cooler}
\end{figure}

Samples have been operated in a Stirling microcooler \cite{AIM}, at
temperatures from 55 to $82 \,\mathrm{K}$. 
%Active cooling does not degrade the SQIF performance, and offers numerous advantages, such as the possibility to set stable temperatures over a large temperature range, quick thermal cycles, compact and portable setups \cite{Caputo:APL_2004,Oppenlander:ASC_2005}. 
To provide transparency to the
\textit{rf} fields, a glass chamber surrounds the microcooler cold head. 
Figure\thinspace\ref{cooler} shows the typical experimental setup.
The sample is  
anchored to the cold head only from one side, in order to keep the focusing ring as much as possible free of
metallic ground plane. On the cold head, an \textit{rf} amplifier is also mounted,
designed for cryogenic applications with a bandwidth $0.04 - 6 \,\mathrm{GHz}$. At room temperature and at the typical bias values, we have measured a noise figure of the amplifier of about 1 dB, in the frequency range $0.1 - 1 \,\mathrm{GHz}$. 
%The amplifier input resistance is $5 \,\mathrm{k\Omega}$. 
The SQIF is coupled to the amplifier through a coplanar line, after which a bias-Tee is placed. The bias-Tee splits the \textit{rf} and the \textit{dc} signals: the \textit{rf}, through a capacitor, goes to
the first stage of the amplifier; the \textit{dc} signal (bias input \& voltage
output), through a conical inductor, goes to the room temperature electronics,
via twisted wires. 
The amplifier output is carried outside the cooler via an SMA feed-through, and detected by a spectrum analyzer. 

The two tone experiments are performed in the following way. The \textit{rf} fields at frequencies $f_1$ and $f_2$ are broadcast. 
The spectrum analyzer is operated in \textsl{Zero Span Mode}, \ie as a narrowband receiver tuned at the central frequency $f_\mathrm{0}$ (equal to the frequency at which the detection of the SQIF spectral voltage output is made, \ie the mixed frequency $f_\mathrm{0} = f_\mathrm{1} - f_\mathrm{2}$), and with
a bandwidth around $f_\mathrm{0}$ set by the Resolution Bandwidth (ResBW). To this mode, there corresponds an analog output proportional to the maximum amplitude of the  signal at $f_\mathrm{0}$. This output voltage is recorded by computer while slowly sweeping either the static
magnetic field $B$ (simultaneously, the $V_\mathrm{dc}(B)$- curve is acquired), or sweeping the static bias current $I_\mathrm{b}$ (simultaneously, the $V_\mathrm{dc}(I_\mathrm{b})$- curve is acquired).

\subsection{Two tone response on the SQIF $V_\mathrm{dc}(B)$- curve}

\begin{figure}[!htb]
\centering
\includegraphics*{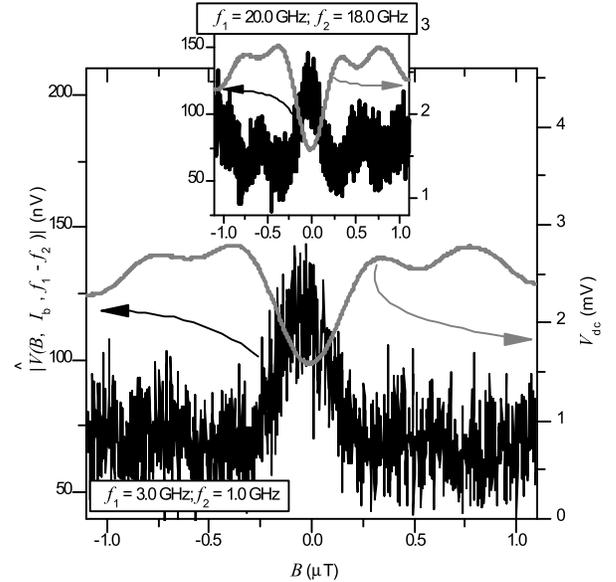}
\caption{The two incident signals are set at $f_{1}=3.0\,\mathrm{GHz}$ and $f_{2}=1.0\,\mathrm{GHz}$. Gray curve, relative to right axis: $V_\mathrm{dc}$ vs. $B\,$; black curve,
rel. to left axis: spectral voltage $\left\vert \widehat{V}\left( B, I_\mathrm{b}, f\right) \right\vert$ detected at $f_0=f_{1}-f_{2} = 2.0\,\mathrm{GHz}$. Inset (axis title and scale as in the figure): $f_{1}=20.0\,\mathrm{GHz}$, $f_{2}=18.0\,\mathrm{GHz}$, and $f_0= 2.0\,\mathrm{GHz}$. For both plots:  $I_\mathrm{b}=20\,\mathrm{\mu A}$ \& $T=\,73\mathrm{K}$; ResBW is $30 \,\mathrm{kHz}$.}
\label{VB}
\end{figure}

We first measure the $%
V_\mathrm{dc}(B)$ dependence of the SQIF at a constant $I_\mathrm{b}$ value, and check the dip symmetry with respect
to zero magnetic field. At $T \approx 73\, \mathrm{K}$ and at $I_\mathrm{b} = 20\, 
\mathrm{\mu A}$, we measure a voltage span $\Delta V \approx 1280\, \mathrm{\mu V}$ and a transfer factor $%
V_\mathrm{B} \approx 5500 \, \mathrm{V/T}$. The SQIF normal resistance is $R = 250\,
\Omega $. Successively, the
\textit{rf} is switched ON: a time harmonic signal which is the superposition of two \textit{rf} signals with frequencies $f_\mathrm{1}$ and $f_\mathrm{2}$ and 
equal amplitudes $b_\mathrm{rf1}$ and $b_\mathrm{rf2}$ is applied to the primary antenna, while the \textit{rf} output of the SQIF is
detected. 
A small amplitude of the incoming signals is required, so that the \textit{rf} signal superimposed to the static field does not
modulated the SQIF working point out of the dip and corrupt the effect; but
it has to be high enough so that the SQIF spectral voltage output is above the noise level
set by the resolution bandwidth (ResBW) of the spectrum analyzer. 
The analog output of the spectrum analyzer is recorded by computer while slowly sweeping the static field $B$
(sweep frequency in the kHz range); simultaneously, the \textit{dc} voltage output $V_\mathrm{dc}$ is
acquired as a function of $B$. In the output spectra, we find the component at $f_0=\mid {f_1-f_2}\mid$, whose amplitude is maximal at the dip bottom and vanishes in the region of zero curvature of the $V_\mathrm{dc}(B)$ curve. Indeed, the amplitude of the second harmonics (the term in ${f_1-f_2}$) is expected to be proportional to the curvature in the $\partial^2_\mathrm B V_\mathrm{dc}\vert_\mathrm{I_b}$. Naturally, the SQIF spectral voltage contains also harmonics at $f_\mathrm{1}$ and at $f_\mathrm{2}$, with the amplitude dependence proportional to the slope       
$\partial_\mathrm B V_\mathrm{dc}\vert_\mathrm{I_b}$, as reported in \cite{Caputo:APL_2006}. We have detected the two tone response of incident fields at frequencies up to 20 GHz (upper limit of our signal generators), keeping the difference frequency $f_\mathrm{0}$ within the bandwidth of the \textit{rf} amplifier. Figure\thinspace\ref{VB} and its inset display the typically measured curves. The incident field is a linear
combination of two signals with frequencies $f_{1}=3.0\,\mathrm{%
GHz}$ and $f_{2}=1.0\,\mathrm{GHz}$, and equal amplitudes (-24 dBm). In the spectral voltage output of the SQIF, we find the  component at the difference frequency $f_\mathrm{0}=2.0\,
\mathrm{GHz}$, and in the Figure is displayed the spectral voltage output $\left\vert \widehat{V}(B,I_\mathrm{b}, f) \right\vert$ detected at the central frequency $f_\mathrm{0}$, with ResBW $=30\,\mathrm{kHz}$, as a function of the sweeping parameter $B$. The bias current was $I_\mathrm{b}= 20 \mathrm{\mu A}$, and $T=\,73\mathrm{K}$. Around $f_0 = 2.0 \,\mathrm{GHz}$, the gain of the amplifier is about $40\, \mathrm{dB}$. The amplitude of the spectral voltage output is maximal around $B=0$, and vanishes as the dc working point approaches the dip slopes; eventually, in other regions of the $V_\mathrm{dc}(B)$-curve with non-zero curvature, the detected spectral voltage rises again from the noise level. In the inset of Fig.\thinspace\ref{VB} is reported the same type of measurement, with the following parameters: the incident fields are at $f_{1}=20.0\,\mathrm{GHz}$ and $f_{2}=18.0\,\mathrm{GHz}$ (amplitudes -15 dBm); the detection is made at $f_0= 2.0\,\mathrm{GHz}$, with $\mathrm{ResBW}=30\,\mathrm{kHz}$.
%; the SQIF bias is $20 \mathrm{\mu A}$, and $T=\,73\mathrm{K}$. 
As one can see, the two tone response is also present for incoming signals with larger frequencies.
Indeed, we have detected the quadratic harmonics over a broad frequency range: for incident fields from a few hundred MHz up to 20 GHz, keeping their distance at various constant values suitable for the bandwidth of our amplifier (within $100 \mathrm{MHz} - 6 \mathrm{GHz}$).  
Qualitatively, no significant changes were observed, over the entire frequency range. A part from the levels of the output signals, which systematically tend to lower at higher frequencies.

\subsection{Two tone response on the SQIF $V_\mathrm{dc}(I_b)$- curve}
\begin{figure}[!tb]
\centering
\includegraphics*{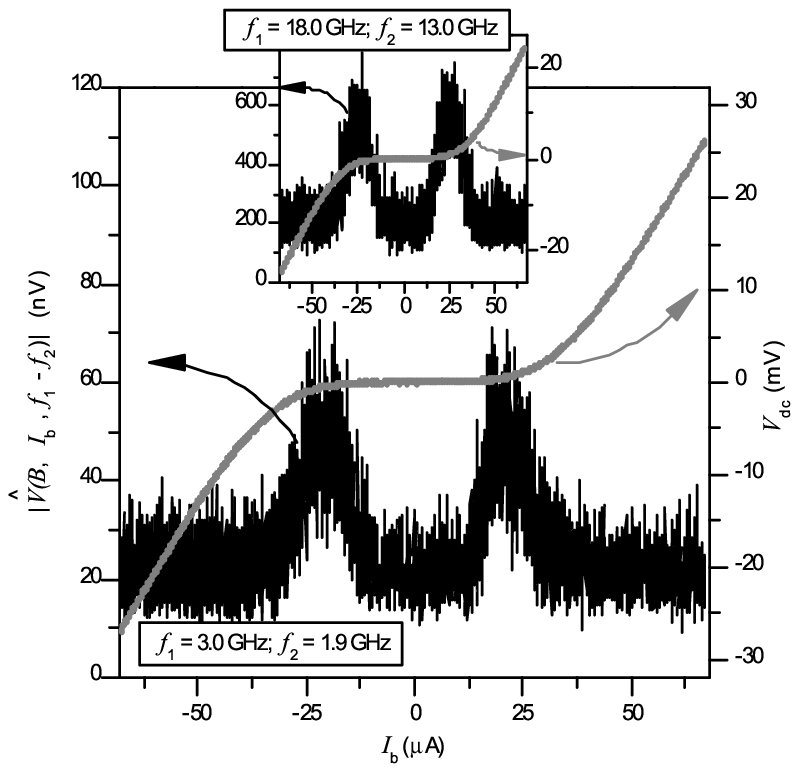}
\caption{The two incident signals are set at $f_{1}=3.0\,\mathrm{GHz}$ and $f_{2}=1.9\,\mathrm{GHz}$. 
Gray curve, relative to right axis: $V_\mathrm{dc}$ vs. $I_\mathrm{b}\,$, at $B = 0$; black curve,
rel. to left axis: spectral voltage $\left\vert \widehat{V}\left( B, I_\mathrm{b}, f\right) \right\vert$ detected at $f_\mathrm{0}= {f_{1}-f_{2}} =1.1\,\mathrm{GHz}$; 
Inset (axis title and scale as in the figure): $f_{1}=18.0\,\mathrm{GHz}$, $f_{2}=13.0\,\mathrm{GHz}$, and 
$f_\mathrm{0}= 5.0\,\mathrm{GHz}$.
For both plots, the ResBW is $3\,\mathrm{kHz}$,  and $T=73.5\,\mathrm{K}$. }
\label{IV}
\end{figure}

At $B=0$, the $V_\mathrm{dc}(I_\mathrm{b})$ curve is swept; the two incident \textit{rf} signals are applied to the primary antenna, and the spectral voltage output of the SQIF sensor is detected. To the applied time harmonic magnetic field, there corresponds naturally a time harmonic electric field, which couples to the SQIF voltage output through the fluctuations induced in the bias current (if we assume the electric field to have a component parallel to the bias line). Thus, the spectral voltage output detected \vs $I_\mathrm{b}$ might also contain fingerprints of the incident signals, and eventually mix them.  Figure\thinspace\ref{IV} and its inset display the typically measured curves. The incident field is a linear
combination of two signals with frequencies $f_\mathrm{1}=3.0\,\mathrm{%
GHz}$ and $f_\mathrm{2}=1.9\,\mathrm{GHz}$, and amplitudes - 44 dBm. In the spectral voltage output of the SQIF, we find the 
quadratic mixing component at $f_\mathrm{0}=f_\mathrm{1}-f_\mathrm{2}=1.1\,
\mathrm{GHz}$, detected with a ResBW of $3\,\mathrm{kHz}$. Also in this case, the amplitude of the voltage output at $f_\mathrm{1}-f_\mathrm{2}$ is found to be maximal around the region of maximal curvature of the $V_\mathrm{dc}(I_\mathrm{b})$- characteristics. 
In the inset of Fig.\thinspace\ref{IV} is reported the same type of measurement, with the following parameters: the incident fields are  generated at $f_{1}=18.0\,\mathrm{GHz}$ and $f_{2}=13.0\,\mathrm{GHz}$, with amplitudes - 14 dBm; the detection is made at $f_0= 5.0\,\mathrm{GHz}$, with $\mathrm{ResBW}=3\,\mathrm{kHz}$. 
In contrast to the case of the $V_\mathrm{dc}(B)$-curve, when sweeping the $V_\mathrm{dc}(I_\mathrm{b})$-curve the spectral voltage output contains a component at $f_\mathrm{0}= 2f_\mathrm{2}-f_\mathrm{1}$, with an amplitude dependence proportional to the third order derivative of the $V_\mathrm{dc}(I_\mathrm{b})$-curve \cite{Cardiff_2006}. 
 
%\subsection{Evidence of electric bias coupling to the electromagnetic fields}
\subsection{Frequency dependence of the two tone response on the $V_\mathrm{dc}(I_b)$- curve}

\begin{figure}[!tb]
\centering
\includegraphics*{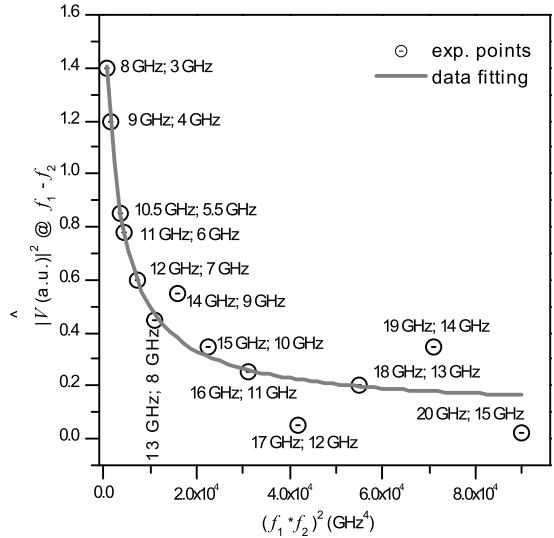}
\caption{Maximum power levels of the spectral component detected at $f_0=f_{1}-f_{2} = 5.0\,\mathrm{GHz}$, for various frequencies of the two incident signals of constant amplitudes (-40 dBm), and plotted (circles) as a function of the squared product of the corresponding frequencies. The power levels were taken in the region of maximum curvature of the $V_\mathrm{dc}(I_\mathrm{b})$-curve. For all points, ResBW is $3\,\mathrm{kHz}$. The data fitting (continuous line) is $\propto (f_1 \cdot f_2)^{-2}$.}
\label{Vvsf}
\end{figure}

We have analyzed the amplitude of the two tone response \vs frequency of the incident fields, keeping constant the difference frequency $f_\mathrm{0}$. Over the whole frequency range, the two fields were broadcast with constant amplitudes equal to -40 dBm. This does not necessary imply that the SQIF-sensor was irradiated with constant energy, since the efficiency of the primary antenna depends on frequency. The two tone response \vs frequency was measured on the $V_\mathrm{dc}(I_b)$- curve. %, and the SQIF spectral voltage was detected at $f_\mathrm{0} = 5 \mathrm{GHz}$. 
As pointed out in the previous section, the presence of the two tone response when sweeping the $V_\mathrm{dc}(I_b)$- curve might be ascribed to the voltage fluctuations induced by the rf electric field via coupling to the bias current.
In Fig.\thinspace\ref{Vvsf}, the y-axis represents, for a given pair of incident fields broadcast at $f_\mathrm{1}$ and $f_\mathrm{2}$, the maximum amplitude of the spectral component detected at the difference frequency $f_\mathrm{0} = 5 \mathrm{GHz}$. Since the directly measured amplitude is in fact the power level [$\propto V^2$] of the spectral component at $f_0$, the data are plotted as a function of the \emph{square} of the product of the two frequencies [$(f_1 \cdot f_2)^2$]. The data fitting, made with a function proportional to $(f_1 \cdot f_2)^{-2}$, well describes the set of experimental points. As can be demonstrated, in the presence of an \textit{rf} electric field which is a linear combination of two fields at $f_1$ and $f_2$, and which couples to the \textit{dc} bias current by adding dispersive fluctuations, the spectral voltage output $V(I_b + i_{rf}(t))$ at $f_1 - f_2$ has a $(f_1 \cdot f_2)^{-1}$-dependence in the prefactor. Thus, in Fig.\thinspace\ref{Vvsf} the observed dependence on frequency of the power levels of the spectral component detected at $f_0$ \vs the squared frequency product $(f_1 \cdot f_2)^2$ is yet another evidence of the electric bias coupling to the applied electromagnetic fields.  
The observed dependence on frequency reflects the $1/\omega$ dependence of the conductivity $\sigma(\omega)$ in a superconductor. The strength of this coupling is comparable to that of the magnetic coupling (induced by the time dependent magnetic flux threading the SQIF loops at the rate of the incident fields).

\section{Conclusions}
We have studied the two tone response of Superconducting Quantum Interference Filters over a frequency range from few MHz to 20 GHz. Incident \textit{rf} fields couple either through the time dependent magnetic flux threading the loops of the interferometer, or through the bias current, so that the response arises both on the $V_\mathrm{dc}(B)$- and $V_\mathrm{dc}(I_b)$-curves. 
The experiments reveal independence of the two tone response on the sweeping parameter, either $B$ or $I_\mathrm{b}$, irrespective of the frequency of the individual tones $f_\mathrm{1}$ and $f_\mathrm{2}$. This is expected if one trusts a determination of the Josephson frequency from an estimation of the noise-free critical current, which gives a value around 40 GHz for our JJ's \cite{IJ_and_SCH}.  
%
\iffalse
\footnote{$f_J \approx 4 \mathrm{GHz}$, as calculated from the \textit{dc} voltage working point set by $B$ and $I_\mathrm{b}$. This value is estimated with an accuracy of about 30\%, which is the typical spread of the grain boundary junction critical currents. In fact, $V_\mathrm{dc}$ is measured for a series of 211 loops and due to the spread of the junction parameters along the array, it might vary correspondingly from loop to loop.},

in the spectral voltage output of the SQIF the two tone response is clearly evident, with the typical amplitude dependence on the sweeping parameter.
\fi
%
When the two tone response arises on the $V_\mathrm{dc}(I_b)$- curve, its voltage amplitude as a function of the frequencies of the incident \textit{rf} fields decays as $1/(f_1 \cdot f_2)$, confirming that the broadcast electromagnetic fields couple to the SQIF not only through the time dependent magnetic flux threading the loops, but also through the voltage fluctuations induced by the bias current fluctuations due to the time dependent electric field.

% or

% Use this command to get the appendices' numbers in "A", "B" instead of the
% default capitalized Roman numerals ("I", "II", etc.).
% However, the capital letter form may result in awkward subsection numbers
% (such as "A-A"). Capitalized Roman numerals are the default.
%\useRomanappendicesfalse
%

\section*{Acknowledgments}
% optional entry into table of contents (if used)
%\addcontentsline{toc}{section}{Acknowledgment}
The authors would like to acknowledge many useful discussions with R.~IJsselsteijn and V.~Schultze.



% http://www.ctan.org/tex-archive/macros/latex/contrib/supported/IEEEtran/bibtex
%\bibliographystyle{IEEEtran.bst}

\begin{thebibliography}{1}

\bibitem{Oppenlander:PRB} J.~Oppenl{\"a}nder, Ch.~H{\"a}ussler, and N.~Schopohl, Phys.\ Rev. B, 
{\bf {63}}, 024511 (2000).

\bibitem{Haussler:JAP2001} Ch.~H{\"a}ussler, J.~Oppenl{\"a}nder, and N.~Schopohl, J. Appl. Phys., {\bf{89}}, 1875 (2001).

\bibitem{Schultze:SST_2003} V.~Schultze, R.~IJsselsteijn, R.~Boucher, H.--G.~Meyer, J.~Oppenl{\"a}nder and {Ch. H{\"a}ussler}, and N.~Schopohl, Supercond. Sci. Technol., {\bf{16}}, 1356 (2003). 

\bibitem{Schultze:SST_2005} V.~Schultze, R.~IJsselsteijn and H.--G.~Meyer, Supercond. Sci. Technol., {\bf{19}}, 411 (2006).

\bibitem{Caputo:APL_2006} P.~Caputo, J.~Tomes, J.~Oppenl{\"a}nder, {Ch. H{\"a}ussler}, A.~Friesch, T.~Tr{\"a}uble, and N.~Schopohl, Appl. Phys. Lett., {\bf{89}}, 062507 (2006).

\bibitem{IPHT_Jena} Institute for Physical High Technology (IPHT), Jena, Germany.

%H.~Kopka and P.~W. Daly, \emph{A Guide to {\LaTeX}}, 3rd~ed.\hskip 1em plus

%%  
\bibitem{Ludwig:IEEE2001} F.~Ludwig, A.~B.~M.~Jansman, D.~Drung, M.~O.~Lindstr{\"o}m, S.~Bechstein, J.~Beyer, J.~Flokstra, and T.~Schurig, IEEE Trans. Appl. Supercond., {\bf{11}}, 1315 (2001). 

\bibitem{AIM} AIM, AEG Infrarot-Module GmbH, Heilbronn, Germany.

\bibitem{Oppenlander:ASC_2005} J.~Oppenl{\"a}nder, {Ch. H{\"a}ussler}, A.~Friesch, J.~Tomes, P.~Caputo, T.~Tr{\"a}uble, and N.~Schopohl, IEEE Trans. Appl. Supercond., {\bf{15}}, 936 (2005).

\bibitem{Caputo:APL_2004} P.~Caputo, J.~Oppenl{\"a}nder, Ch. H{\"a}ussler, J.~Tomes, A.~Friesch, T.~Tr{\"a}uble, and N.~Schopohl, Appl. Phys. Lett., {\bf{85}}, 1389 (2004).


\bibitem{Cardiff_2006} P.~Caputo, J.~Tomes, J.~Oppenl{\"a}nder, {Ch. H{\"a}ussler}, A.~Friesch, T.~Tr{\"a}uble, and N.~Schopohl, accepted for publication in J. Supercond. (as proceeding of HTSHFF Symposium, Cardiff 2006)

\bibitem{IJ_and_SCH} R.~IJsselsteijn and V.~Schultze, private communication.

\end{thebibliography}

\vfill

\end{document}